\documentclass[a4paper,11pt]{article}
\pdfoutput=1

\usepackage{jheppub}
\usepackage{amsmath,amssymb}
\usepackage{tikz-cd}
\usepackage{booktabs}
\usepackage[T1]{fontenc}

\title{\texorpdfstring{$A_{\infty}$}{A_infinity} action of open $\mathcal{N} = 2$ superstring field theory}

\author[a]{Xianghang Zhang}

\affiliation[a]{Graduate School of Mathematics, Nagoya University,\\ Furocho 1, Nagoya 464{-}8602, Japan}

\emailAdd{zhang.xianghang.g5@math.nagoya-u.ac.jp}
\abstract{We formulate a string field theory for open $\mathcal{N}=2$ strings with an $A_\infty$ algebra structure. Starting from the BRST cohomology relative to the $U(1)$ anti-ghost zero-mode, we generalize [arXiv:1312.2948] and constructed all interacting vertices recursively and without singularity. We also show that our string field theory reproduces the correct perturbative S-matrix.}

\begin{document}

\maketitle
\flushbottom

\section{Introduction}
\label{sec:intro}

String perturbation theory is a theory of on-shell amplitudes defined by integrals of conformal field theory correlators over the (super)moduli space of the worldsheet.
Its second quantization, known as string field theory is formulated via an action principle that reproduces the perturbative amplitudes.
It thus amounts to a decomposition of the moduli space according to the Feynman rule~\cite{Doubek2020}.

This geometric approach has been clarified in particular for bosonic strings.
Witten's cubic open string field theory~\cite{Witten1986Non} is a prominent example.
It is an elegant choice that the cubic vertex in the action, after Feynman expansion, covers the entire moduli space with an arbitrary number of vertex operators insertions~\cite{Zwiebach1991}.
In his monumental work~\cite{Zwiebach1993}, Zwiebach identified the homotopy algebra structures ($A_{\infty}$ and $L_{\infty}$ for open and closed strings, respectively) in the bosonic string field theory action and derived the minimal area principle that determines all vertices for bosonic closed strings.
From a field-theoretic perspective, the homotopy algebra encodes the gauge invariance of the action.
In Witten's cubic action, the algebraic structure is simply that of a graded associative algebra since the star product is associative and there is no need for higher order terms.
For a general open string field theory, we consider an $A_{\infty}$ algebra, which is a graded associative algebra generalized up to homotopy.

Such a geometric picture, however, becomes significantly more subtle in the context of superstrings.
One alternative yet successful approach to superstring field theory, due to Berkovits, is an analogue of the Wess-Zumino-Witten action~\cite{BerkovitsSup1995,Berkovits2000}.
The action is presented in closed form and yields various analytic solutions~\cite{Ohmori2002,Kluson2001,Kluson2002,Lechtenfeld2002,Erler2013}.
However, in Berkovits' formulation the geometric picture is largely absent.
This is partly due to the bosonization of the $\beta\gamma$ superconformal ghosts, which introduces an unphysical degree of freedom, resulting in a ``large'' Hilbert space.
The lack of geometric understanding also complicates quantization~\cite{Berkovits2012}.
A deeper understanding of how the superstring field action encodes supermoduli space data is necessary.

In the past decade another route with a closer connection to the moduli space has been pursued.
The conventional method of computing superstring amplitudes is to first integrate over the fermionic part of the moduli space, leaving picture-changing insertions on the worldsheet (see, e.g.,~\cite{Witten2012}).
One natural modification of Witten's cubic string field theory is to dress the star product with a picture-changing operator at the midpoint~\cite{Witten1986Int}, but this naive idea suffers from divergent contact terms at higher-point interactions~\cite{Wendt1989}.
Modifications of this idea have appeared~\cite{Lechtenfeld1988,Preitschopf1990,Arefeva1990,Berkovits2009,Kroyter2011}, but all require changing the kinetic term, yielding complicated propagators.
However, with the help of $A_{\infty}$ relations, a consistent way of smearing and assigning picture-changing opeartors to the vertices was developed by Erler, Konopka and Sachs~\cite{Erler2014} and later extended to the Ramond sector~\cite{Erler2016}.
Though not directly built from the supermoduli space, a geometric interpretation using vertical integration~\cite{Sen2015} was formulated in~\cite{Erler2017}.
A key advantage of the homotopy algebra approach is its compatibility with Batalin-Vilkovisky quantization~\cite{Zwiebach1993}; however quantum-level developments remain limited.

On the other hand, the term ``superstring'' conventionally refers to the five theories possessing $\mathcal{N} = 1$ superconformal symmetry on the worldsheet.
However, though often overlooked, under the requirement of a positive critical dimension, it is also possible to impose $\mathcal{N} = 2$ superconformal symmetry on the worldsheet theory, which has a curious property that all massive excited states are absent.
Given the success of the complete homotopy-algebraic formulation of $\mathcal{N} = 1$ superstring field theory, it is natural to ask whether similar constructions apply to $\mathcal{N} = 2$ strings.

Our understanding of the spacetime geometry aspects of $\mathcal{N} = 2$ strings was greatly improved by the work of Ooguri and Vafa~\cite{Ooguri1990,Ooguri1991}, which revealed their connection to the self-dual sector of Yang-Mills and Einstein theory in Euclidean or split signature\footnote{In Euclidean signature, the on-shell condition becomes problematic, hence we focus on split signature with two timelike dimensions}.
While at first glance the critical dimensionality does not seem to promise interesting phenomenology, connections have been suggested from self-dual theory to maximally helicity-violating QCD amplitudes~\cite{Bardeen1996,Chalmers1996}, to super Yang-Mills~\cite{Neitzke2004}, to F-theory~\cite{Vafa1996} and to integrability~\cite{Hamanaka2023}.
Moreover, there have also been attempts to include spacetime fermions~\cite{Ketov1995twist,Siegel1992}.

Another motivation for studying $\mathcal{N} = 2$ string field theory comes from the attempt to understand the underlying geometry of Berkovits' formulation.
In his original proposal, Berkovits started by  enlarging and twisting the $\mathcal{N} = 2$ superconformal algebra and obtained an interacting theory of string field.
The action of $\mathcal{N} = 1$ strings can then be obtained by embedding it into a $\mathcal{N} = 2$ superconformal algebra~\cite{Berkovits1994,BerkovitsTop1995} and consequently, the reason why Berkovits's formulation works so well is likely to be unraveled from the geometric aspect of the $\mathcal{N} = 2$ moduli space.

This paper is a first step towards the goal by generalizing the homotopy-algebraic method to the $\mathcal{N} = 2$ case.
The paper is organized as follows.
We review generalities of open $\mathcal{N} = 2$ string theory in section~\ref{sec:rel}.
Section~\ref{sec:con} is the main part of this article, where we generalize the homotopy-algebraic construction of Erler, Konopka and Sachs to the case involving two picture-changing operators are present.
In section~\ref{sec:amp}, we prove that the action derived in this framework reproduces worldsheet scattering amplitudes — a direct extension of previous work~\cite{Konopka2015} in the $\mathcal{N} = 1$ case.
We conclude with a discussion of implications and future directions.

\section{Free theory}
\label{sec:rel}

In this section, we review some facts about $\mathcal{N} = 2$ string theory and write down the free string field theory action using relative cohomology.
For a detailed exposition on the worldsheet theory, see for example~\cite{Bischoff1997Int}.
The flat Kleinian target spacetime $\mathbb{R}^{2,2}$, where critical $\mathcal{N} = 2$ strings propagate, is endowed with a complex structure of $\mathbb{C}^{1,1}$.
It is a curious fact that the theory requires a chosen complex structure, thus breaking the $SO(2,2)$ symmetry into $U(1,1)$.
A method of restoring the symmetry was developed in~\cite{Bischoff1997Res}, essentially averaging over all complex structures, but this will not be the main topic of this article.
On the worldsheet, there is a theory of $\mathcal{N} = 2$ supergravity $(h,\,\chi,\,A)$ containing the metric, the gravitino fields and a $U(1)$ gauge field on a two-dimensional Riemann surface with boundaries, coupled to two chiral $\mathcal{N} = 2$ massless matter multiplets $(Z^{ij}, \psi^{ij})$, where $i \in \{+, -\}$ denotes the spacetime complex conjugations on $\mathbb{C}^{1,1}$, and $j \in \{+, -\}$ is the light-cone index in the usual sense:
\begin{equation}
  \begin{split}
    Z^{\pm j} &= Z^{1j} \pm iZ^{2j}, \\
    Z^{1\pm} &= X^{0} \pm X^{2}, \\
    Z^{2\pm} &= X^{1} \pm X^{3}, \\
  \end{split}
\end{equation}
where $(X^{0}, X^{1}, X^{2}, X^{3})$ are coordinates on the real space $\mathbb{R}^{2,2}$.
In this notation the scalar product of two $\mathbb{R}^{2,2}$ vectors $k$ and $p$ can be expressed as
\begin{equation}
\begin{split}
  k \cdot p &= \frac{1}{2}\Big( k^{+j}p^{-}{}_{j} + k^{-j}p^{-}{}_{j} \Big) \\
            &= \frac{1}{4}\Big( k^{++}p^{--} + k^{+-}p^{-+} + k^{-+}p^{+-} + k^{--}p^{++} \Big) \\
            &= k^{0}p^{0} + k^{1}p^{1} - k^{2}p^{2} - k^{3}p^{3} ,
\end{split}
\end{equation}
where the summation over repeated index $j$ is assumed and the second index is lower using the light-cone metric in the first line, such that
\begin{equation}
k^{i}\cdot p^{i'} \equiv k^{ij} p^{i'}{}_{j} = \frac{1}{2} \left( k^{i+} p^{i'-} + k^{i-} p^{i'+} \right).
\end{equation}
Note that $Z$ is a complex boson and $\psi$ is a Dirac spinor on the worldsheet when spacetime is complex.
We denote the spinor index by a lower index of the spin orientation
\begin{equation}
  \psi^{ij} =
   \begin{pmatrix}
    \psi^{ij}_{\uparrow}\\
    \psi^{ij}_{\downarrow}
  \end{pmatrix}
\end{equation}
The worldsheet action after gauge fixing $(h^{\alpha\beta}=\delta^{\alpha\beta}, \chi=0, A=0)$ is given by~\cite{Brink1977}
\begin{equation}
  S = -\frac{1}{4\pi} \int \mathrm{d}^{2}z (\partial Z^{-j}\bar{\partial}Z^{+}{}_{j} + \bar{\partial} Z^{-j}\partial Z^{+}{}_{j} + \psi_{\uparrow}^{-j}\overleftrightarrow{\partial}\psi^{+}_{\uparrow}{}_{j} + \psi_{\downarrow}^{-j}\overleftrightarrow{\bar{\partial}}\psi^{+}_{\downarrow}{}_{j}),
\end{equation}
where we have set the Regge slope to $1$.

Gauge fixing requires introducing the ghosts $(c, \tilde{c}, \gamma^{\pm})$ for the diffeomorphism, $U(1)$ gauge symmetry and $\mathcal{N}=2$ local supersymmetry, respectively.
The Jacobian of the change of path integral variables is compensated by the ghost action
\begin{equation}
  S_{gh} = \frac{1}{\pi}\int\left( b\bar{\partial}c + \tilde{b}\bar{\partial}\tilde{c} + \beta^{-}\bar{\partial}\gamma^{+} + \beta^{+}\bar{\partial}\gamma^{-} \right),
\end{equation}
which additionally consists of anti-ghosts $(b, \tilde{b}, \beta^{\mp})$.
The ghosts obey the following OPEs:
\begin{equation}
b(z)c(w) \sim \frac{1}{z-w},
\end{equation}
\begin{equation}
\beta^{\pm}(z)\gamma^{\mp}(w) \sim -\frac{1}{z-w},
\end{equation}
and
\begin{equation}
\tilde{b}(z)\tilde{c}(w) \sim \frac{1}{z-w}.
\end{equation}
The theory contains a $\mathcal{N} = 2$ superconformal algebra generated by the currents:
\begin{equation}
  \begin{split}
    T &= T_{m} +\partial c b+c \partial b+\partial\tilde{c} \tilde{b}-\frac{3} {2} ( \partial\gamma^{-} \beta^{+}+\partial\gamma^{+} \beta^{-} )-\frac{1} {2} ( \gamma^{-} \partial\beta^{+}+\gamma^{+} \partial\beta^{-} ), \\
    G^{\pm} &= G^{\pm}_{m}  - 4 \gamma^{\pm} b \mp 4 \partial\gamma^{\pm} \tilde{b} \mp 2 \gamma^{\pm} \partial\tilde{b}+\frac{3}{2} \partial c \beta^{\pm}+c \partial\beta^{\pm} \mp\tilde{c} \beta^{\pm}, \\
    J &= J_{m} + \partial(\tilde{c}b) + \gamma^{+}\beta^{-} - \gamma^{-}\beta^{+},
  \end{split}
\end{equation}
where the matter part generators $T_{m}$, $G_{m}$ and $J_{m}$ are not written down explicitly.
The BRST charge $Q$ of this worldsheet theory is  defined by the current
\begin{equation}
Q = \oint \frac{\mathrm{d}z}{2\pi i} J_{B}(z),
\end{equation}
where
\begin{equation}
  \begin{split}
    J_{B} = &c T+\gamma^{+} G^{-}+\gamma^{-} G^{+}+\tilde{c} J+c \partial c b+c \partial\tilde{c} b \\
            &-4 \gamma^{+} \gamma^{-} b+2 \partial\gamma^{-} \gamma^{+} \tilde{b}-2 \partial\gamma^{+} \gamma^{-} \tilde{b} \\
            &+\frac{3}{4} \partial c ( \gamma^{+} \beta^{-}+\gamma^{-} \beta^{+} )-\frac{3}{4} c ( \partial\gamma^{+} \beta^{-}+\partial\gamma^{-} \beta^{+} ) \\
            &+\frac{1}{4} c ( \gamma^{+} \partial\beta^{-}+\gamma^{-} \partial\beta^{+} )+\tilde{c} ( \gamma^{+} \beta^{-}-\partial\gamma^{-} \beta^{+} ). \\
  \end{split}
\end{equation}

The commuting ghost zero modes of $\beta^{\mp}$ and $\gamma^{\pm}$ lead to an infinite degeneracy of the energy vacua, known as the \textit{picture number degeneracy}\footnote{See for example~\cite{Witten2012,Witten2019}.}.
To address this issue, it is convenient to bosonize the superconformal ghosts as
\begin{equation}
  \begin{split}
    \gamma^{\pm} &= \eta^{\pm} e^{\varphi^{\pm}}, \\
    \beta^{\mp} &= e^{-\varphi^{\pm}} \partial\xi^{\mp},
  \end{split}
\end{equation}
and
\begin{equation}
  \begin{split}
    \delta( \gamma^{\pm} ) &= e^{-\varphi^{\pm}}, \\
    \delta( \beta^{\mp} ) &= e^{\varphi^{\pm}},
  \end{split}
\end{equation}
with the OPEs:
\begin{equation}
  \begin{split}
    \varphi^{\pm} ( z ) \varphi^{\pm} ( w ) &\sim -\log ( z-w ), \\
    \xi^{\mp} ( z ) \eta^{\pm} ( w ) &\sim \frac{1} {z-w}.
  \end{split}
\end{equation}
The bosonized fields $\xi^{\pm}$ enter the bosonization only through $\partial \xi^{\pm}$.
As a result, the Hilbert space after bosonization contains unphysical degrees of freedom—namely the zero modes of $\xi^{\pm}(z)$, denoted by
\begin{equation}
\xi^{\pm} \equiv \frac{1}{2\pi i }\oint \frac{\mathrm{d}z}{z} \xi^{\pm}(z).
\end{equation}
Since we consider the open string theory only, doubling trick will be used throughout the paper.
This ``large'' Hilbert space is the starting point of the Berkovits formulation in the $\mathcal{N} = 1$ case~\cite{BerkovitsSup1995}.
By requiring the exclusion of $\xi^{\pm}$ zero modes, we get the physical condition on the large Hilbert space
\begin{equation}
  \mathcal{H} \cong \{\Psi \in \mathcal{H}_{L} | \eta^{+}\Psi = \eta^{-}\Psi = 0\},
\end{equation}
where $\eta^{\pm}$ is the zero mode of $\eta^{\pm}(z)$.
$\mathcal{H}$ can be identified as a subspace of the large Hilbert space
\begin{equation}
\mathcal{H} \subset \mathcal{H}_{L} \cong \mathcal{H}\,\oplus\,\xi^{+}\mathcal{H}\, \oplus\, \xi^{-}\mathcal{H}\, \oplus\, \xi^{+}\xi^{-}\mathcal{H},
\end{equation}
on which the symplectic form $\omega_{L}$ is related to that on the small Hilbert space via
\begin{equation}
\omega = \omega_{L}(\xi^{+}\xi^{-}\otimes \mathbb{I}).
\end{equation}
In the following, $\xi^{\pm}$ will be used in the intermediate steps of the construction, but the physical condition will be used to fix the ambiguity, yielding an action in the small Hilbert space.

The Hilbert space is graded by the picture numbers $(\pi^{+}, \pi^{-})$ summarized in table~\ref{tab:qnumber}.
\begin{table}
  \centering
\begin{tabular}{ccccccccccccc}
\toprule
 & $b$ & $c$ & $\tilde{b}$ & $\tilde{c}$ & $\beta^\pm$ & $\gamma^\pm$ & $\xi^+$ & $\xi^-$ & $\eta^+$ & $\eta^-$ & $\exp{(l\phi^{+})}$ & $\exp{(l\phi^{-})}$ \\
\midrule
$h$ & $2$ & $-1$ & $1$ & $0$ & $3/2$ & $-1/2$ & $0$ & $0$ & $1$ & $1$ & $-\frac{l(l+2)}{2}$ & $-\frac{l(l+2)}{2}$ \\
\midrule
$N_{\mathrm{gh}}$ & $-1$ & $1$ & $-1$ & $1$ & $-1$ & $1$ & $-1$ & $-1$ & $1$ & $1$ & $0$ & $0$ \\
\midrule
$\pi^+$ & $0$ & $0$ & $0$ & $0$ & $0$ & $0$ & $1$ & $0$ & $-1$ & $0$ & $l$ & $0$\\
\midrule
$\pi^-$ & $0$ & $0$ & $0$ & $0$ & $0$ & $0$ & $0$ & $1$ & $0$ & $-1$ & $0$ & $l$ \\
\bottomrule
\end{tabular}
\caption{The conformal weight $h$, the ghost number $N_{\mathrm{gh}}$ and the picture number $\pi^{+}$, $\pi^{-}$ of various operators.\label{tab:qnumber}}
\end{table}
Vertex operators in different pictures can be mapped by the picture changing operators\footnote{In contrast to the $\mathcal{N} = 1$ string, there are no local ``picture-lowering operators'' $Y(z)$ that invert $X(z)$ in a sense that $XYX = X$~\cite{Bischoff1995}, hence ruling out constructions of~\cite{Lechtenfeld1988,Preitschopf1990,Arefeva1990,Kroyter2011}.} (PCO)
\begin{equation}
X^{\pm}(z) = \left\{ Q, \xi^{\pm}(z) \right\}.
\end{equation}
We can then compute the OPE
\begin{equation}
  \begin{split}
    X^{\pm} ( z ) X^{\pm} ( w ) &\sim 0, \\
    X^{\pm} ( z ) X^{\mp} ( w ) &\sim \{Q, \mathrm{singular}\}.
  \end{split}
\end{equation}
There is one more ingredient in the $\mathcal{N} = 2$ theory: the $U(1)$ gauge field $A(z)$.
Although locally one can always fix the gauge $A(z) = 0$, the $U(1)$ moduli must be taken into account when there are punctures or non-trivial homology classes on the worldsheet.
One can shift the holonomy around each puncture using the superconformal generator $J$.
Define the spectral flow operator
\begin{equation}
  I(\theta, z)=\exp\left(\theta\int_{z_0}^{z}\mathrm{d}wJ(w)\right)
\end{equation}
where $z_{0}$ is a reference point common for all $I$ insertions.
The Ramond sector vertices can thus be obtained from Neveu-Schwarz vertices by applying the spectral flow operator, and the fact that $I$ is invertible gives rise to isomorphism between NS and R sectors.
Moreover, since the derivative of $I$ is BRST-exact
\begin{equation}
\partial I(\theta, z) = \theta\{Q, \tilde{b}(z)I(\theta, z)\},
\end{equation}
we can then collect all the puncture twists together without affecting the on-shell amplitudes.
The total twist on the worldsheet is then classified by an additional instanton number $n_{I}$ corresponding to the Chern number of the $U(1)$ bundle (for details on the spectral flow, see~\cite{Bischoff1997Int} and references therein) and only NS vertices appear in the amplitudes.
For this reason, we will only construct the vertices for string fields in the NS sector and claim that they encode all dynamics.

The $N$-point tree-level scattering amplitude of instanton number $n_{I}$ is
\begin{equation}\label{eqn:amp}
  \begin{split}
    A^{n_{I}}_{g, 1, \dots, N} &= \prod_{i=1}^{N-1}\int \mathrm{d}M_{i} \prod_{j=1}^{N - 3}\int \mathrm{d}m_{j} \\
    \times&\langle  ( \omega_{i}, \tilde{b} )(\mu_{j}, b)(X^{-})^{N-2+n_{I}} (X^{+})^{N-2-n_{I}} I^{n_{I}} V_{1}(z_{1}) \dots V_{N}(z_{N})\rangle
  \end{split}
\end{equation}
where $M_{i}$ are the $U(1)$ moduli, $m_{j}$ are the complex moduli, $\omega_{i}$ are tangent to $M_{i}$ and $\mu_{j}$ are the Beltrami differentials tangent to $m_{j}$.
Note that all the $X^{\pm}$, $I$ and $\tilde{c}$ insertions are at arbitrary positons as long as there is no singularity.
The relevant amplitude here for classic open string field theory is the tree-level amplitude at zero instanton number $n_{I} = 0$~\cite{BerkovitsSup1995}.
Similar to bosonic string theory, the moduli integration associated to each punctures can be contracted with the vertex operators.
For $m_{j}$ moduli, this results in the so-called integrated vertex operators~\cite{Witten2012}.
Furthermore, for the $(N-1)$ gauge moduli which should be understood as representing the holonomies around every puncture but one base point, the integration amounts to insertions of $\tilde{b}$ and $J$ around these punctures~\cite{Ketov1995measure,Lechtenfeld1996}
\begin{equation}
\oint \mathrm{d}w' \delta{J(w')} \oint \mathrm{d}w \tilde{b}(w) V_{i}(z_{i}),
\end{equation}
which essentially selects out the $U(1)-$neutral part and projects out one $\tilde{c}$ ghost degenerate part of the string field.
These insertions can be reduced by imposing constraint on the string field, as commented in~\cite{Jnemann1999}.
Note that there is only one zero-mode $\tilde{c}_{0}$ in the $\tilde{b}\tilde{c}$ CFT, therefore, the energy vacuum is doubly degenerate as $\{|0\rangle, \tilde{c}_0|0\rangle\}$ and only the states created from the last vacuum survive the $\tilde{b}$ integrations.
We may define the string field as
\begin{equation}
\Psi(z) = \oint \mathrm{d}w' \delta{J(w')} \oint \mathrm{d}w \tilde{b}(w) V(z),
\end{equation}
where $V(z)$ is the standard vertex operator for computing the amplitudes.
In other words, on a off-shell state $\Psi$ we put the constraint
\begin{equation}
  J_{0}\Psi = \tilde{b}_{0} \Psi = 0,
\end{equation}
which yields a well-defined relative BRST-cohomolgy since $ \{ Q, \tilde{b}_{0}\} = J_{0}$.
Imposing the vanishing of $\tilde{b}_{0}$ mode removes the $\tilde{c}_{0}$ mode dependence of the vertex operator.
The resulting amplitude will be of a form similar to the superstrings
\begin{equation}
  \begin{split}
    A^{n_{I}}_{g, 1, \dots, N} = \prod_{j=1}^{N - 3}\int \mathrm{d}m_{j} \langle (\mu_{j}, b)(X^{-})^{N-2+n_{I}} (X^{+})^{N-2-n_{I}} I^{n_{I}} \tilde{c}_{0} \Psi_{1}(z_{1}) \dots \Psi_{N}(z_{N})\rangle,
  \end{split}
\end{equation}
except for the extra $\tilde{c}$ insertion.
We comment that the condition we impose here is analogous to the level-matching condition $(L_{0} - \bar{L}_{0})\Psi = (b_{0} - \bar{b}_{0})\Psi = 0$ for closed string field theory, which should be viewed as a constraint on the string field configuration, whereas the condition $b_{0}\Psi = 0$ that often appears in the literature is a gauge choice known as Siegel's gauge.
The ghost zero mode should be included in the symplectic form
\begin{equation}
  \omega(A, B) = (-1)^{\mathrm{deg}(A)}\langle A, B \tilde{c}_{0}\rangle \equiv (-1)^{\mathrm{deg}(A)}\left\langle I\circ A(0) B(0) \oint_{|z|=1} \frac{\mathrm{d}z}{z} \tilde{c}(z)\right\rangle
\end{equation}
to saturate the $U(1)$ ghost zero mode.
The non-interacting string field action can then be written as
\begin{equation}\label{eqn:free}
  S[\Psi] = \frac{1}{2}\omega(\Psi, Q\Psi),
\end{equation}
and from now on, we denote the space of picture $(-1, -1)$ string fields $\Psi$ satisfying $\tilde{b}_{0}\Psi = J_{0}\Psi = 0$ as $\mathcal{H}$, and the corresponding large Hilbert space $\mathcal{H}_{L}$.

A typical vertex operator in the canonical picture $(-1, -1)$ that will appear in the amplitudes is of the form $V = \tilde{c}c \exp (-i\phi^{+}-i\phi^{-}) V_{m}$, where $V_{m}$ is a matter primary field.
To make up a non-vanishing correlator, the fields must carry total picture number $(-2, -2)$ and thus the quadratic term requires no picture modification.
However, as we see in the following, for a $\Psi^{n}$ term in the action, $(n-2)$ PCO insertions for $X^{+}$ and $X^{-}$ need to be included in the action.

\section{Interacting theory}
\label{sec:con}

In the free theory~(\ref{eqn:free}), only the kinetic term appears, representing an open $\mathcal{N}  = 2$ string propagating without interaction.
An interacting action must include higher order terms representing interactions.
We already know the way to do it gauge invariantly is to exploit the $A_{\infty}$ algebra structure~\cite{Gaberdiel1997}.
For bosonic string, it is related to the decomposition of the moduli space of Riemann surfaces with boundaries, and an elegant construction is Witten's star product.
For a string field $\Psi$, the action is
\begin{equation}\label{eqn:wstar}
S_{\mathrm{bos,\, open}}[\Psi] = \frac{1}{2}\omega(\Psi, Q\Psi) +\frac{g}{3}\omega(\Psi, \Psi \ast \Psi),
\end{equation}
where $\omega$ is the symplectic bilinear form defined via the BPZ inner product $\langle \cdot, \cdot\rangle$, $Q$ is the BRST charge, $g$ is the open string coupling constant, and $\ast$ is the Witten star product.
More subtleties appear when it is generalized to $\mathcal{N} = 1$ and $\mathcal{N} = 2$ supermoduli~\cite{Juro2013}.
Luckily, we are able to avoid this, just like in standard string theory textbook, by integrating out the fermionic variables and assigning the picture-changing operators to each vertex.
Indeed, Witten's original proposal in~\cite{Witten1986Int} is to dress the bosonic product in~(\ref{eqn:wstar}) with a PCO inserted at the midpoint.
\begin{equation}\label{eqn:wstarsup}
S_{\mathrm{sup,\, open}}[\Psi] = \frac{1}{2}\omega(\Psi, Q\Psi) +\frac{g}{3}\omega(\Psi, X(\mathrm{mid})\Psi \ast \Psi),
\end{equation}
However, the singular OPE of $X$ with itself
\begin{equation}
X(\mathrm{mid})X(\mathrm{mid}) \sim \infty
\end{equation}
makes the higher terms divergent.
Note that $\partial X$ is BRST-exact and thus the PCO insertion can be moved arbitrarily on the worldsheet without affecting on-shell amplitudes.
It has been known that this issue of action~(\ref{eqn:wstarsup}) can be resolved by moving the PCO insertion away from the midpoint.
As a result of moving, the new string product is no longer associative, and higher terms must be included.
But there has not been a systematic way to write all higher terms compatible with the $A_{\infty}$ relations until about a decade ago. In~\cite{Erler2014} the ghost mode $\oint \frac{\mathrm{d}z}{2\pi i} \frac{\xi(z)}{z} \equiv \xi$ whose BRST variation gives resolved PCO insertions
\begin{equation}
\{Q, \xi\} = \oint \frac{\mathrm{d}z}{2\pi i} \frac{X(z)}{z} \equiv X,
\end{equation}
is used in the intermediate steps, whereas the physical condition that eliminate the $\xi$ zero mode in the final result is used to fix the ambiguity in solving the $A_{\infty}$ relations.

In $\mathcal{N} = 2$ string theory, the divergence of Witten's theory is more apparent even at three-point level: both $X^{+}$ and $X^{-}$ must be inserted at the midpoint.
The cubic term
\begin{equation}
S_{\mathrm{cubic}}[\Psi] \sim X^{+}(\mathrm{mid})X^{-}(\mathrm{mid}) \Psi^{3} \sim \infty,
\end{equation}
which needs to be resolved similarly.
It is the purpose of this section to argue that such a method can be generalized to the case of open $\mathcal{N}=2$ strings.
To begin with, we briefly review the construction in~\cite{Erler2014} with a slightly more general setup.

\subsection{Algebraic preliminaries}
A generic action of field $\Psi$ living in a graded vector space $\mathcal{H}$ with cyclic $A_{\infty}$ structure can be written as
\begin{equation}
  S[\Psi] = \frac{1}{2}\omega(\Psi, Q\Psi) + \sum_{n=2}^{\infty}\frac{g^{n-1}}{n+1}\omega(\Psi, M_{n}(\Psi^{\otimes n})),
\end{equation}
where $M_{n}:\mathcal{H}^{\otimes n}\to\mathcal{H}$ are called $n$-products.
It forms a \textit{cyclic $A_{\infty}$ algebra} if the cyclicity
\begin{equation}
  \begin{split}
    &\omega( M_{n} ( \Psi_{1},..., \Psi_{n} ), \Psi_{n+1} )\\
    &= (-1 )^{\mathrm{deg} ( \Psi_{1} ) ( \mathrm{deg} ( \Psi_{2} )+...+\mathrm{deg} ( \Psi_{n+1} ) )} \, \times\omega( M_{n} ( \Psi_{2},..., \Psi_{n+1} ), \Psi_{1} ),
  \end{split}
\end{equation}
and $A_{\infty}$ relations
\begin{equation}
  \begin{split}
    Q^{2}\Psi &= 0, \\
    QM_{2}(\Psi_{1}, \Psi_{2}) + M_{2}(Q\Psi_{1}, \Psi_{2}) &+ (-1)^{\mathrm{deg}(\Psi_{1})}M_{2}(\Psi_{1}, Q\Psi_{2}) = 0, \\
    Q M_{3} ( \Psi_{1}, \Psi_{2}, \Psi_{3} )+M_{2} ( M_{2} ( \Psi_{1}, \Psi_{2} ), \Psi_{3} )&+(-1 )^{\operatorname{d e g} ( \Psi_{1} )} M_{2} ( \Psi_{1}, M_{2} ( \Psi_{2}, \Psi_{3} ) ) \\
    M_{3} ( Q \Psi_{1}, \Psi_{2}, \Psi_{3} )+(-1 )^{\operatorname{d e g} ( \Psi_{1} )} M_{3} ( \Psi_{1}, Q \Psi_{2}, \Psi_{3} )&+(-1 )^{\operatorname{d e g} ( \Psi_{1} )+\operatorname{d e g} ( \Psi_{2} )} M_{3} ( \Psi_{1}, \Psi_{2}, Q \Psi_{3} )=0 , \\
    &\cdots
  \end{split}
\end{equation}
are satisfied.
We will sometimes denote $Q$ as $M_{0}$ to get align with the notations.
This action is invariant under the gauge transformation
\begin{equation}
\delta \Psi = \sum_{n=1}^{\infty}\sum_{k=0}^{n-1}M_{n}(\Psi^{n-k-1}\otimes \Lambda \otimes \Psi^{k}),
\end{equation}
where $\Lambda$ is the gauge parameter.
For open string field theory, the grading on $\mathcal{H}$ is defined as the shifted Grassmann parity,
\begin{equation}
  \mathrm{deg}(\Psi) = \mathrm{gr}(\Psi) + 1.
\end{equation}

The $A_{\infty}$ relations can be conveniently expressed if we use the coalgebra notation briefly review here.
Consider the tensor algebra of it the space of string fields $\mathcal{H}$
\begin{equation}
  T\mathcal{H} = \oplus_{n=0}^{\infty}\mathcal{H}^{\otimes n}.
\end{equation}
Any $n$-product $b_{n}: \mathcal{H}^{\otimes n} \to \mathcal{H}$ can be extended to a coderivation on the tensor coalgebra $\mathbf{b}_{n}: T\mathcal{H} \to T\mathcal{H} $, denoted by bold font letters, that is
\begin{equation}
  \mathbf{b}_{n} = \sum_{N=n}^{\infty}\sum_{k = 0}^{N - n}\mathbb{I}^{\otimes N-n-k}\otimes b_{n} \otimes \mathbb{I}^{\otimes k} \circ P_{N},
\end{equation}
where $P_{N}$ is the projector to the $N$-fold tensor space.
Given two coderivations $\mathbf{b}_{n}$ and $\mathbf{c}_{m}$, their commutator is another coderivation derived from a $(n+m-1)$-product
\begin{equation}
[ b_{m}, c_{n} ] \equiv b_{m} \circ \sum_{k=0}^{m-1} \mathbb{I}^{\otimes m-1-k} \otimes c_{n} \otimes\mathbb{I}^{\otimes k}-(-1 )^{\mathrm{d e g} ( b_{m} ) \mathrm{d e g} ( c_{n} )} c_{n} \circ \sum_{k=0}^{n-1} \mathbb{I}^{\otimes n-1-k} \otimes b_{m} \otimes\mathbb{I}^{\otimes k}.
\end{equation}
With the coalgebra notation, the $A_{\infty}$ relations can be compactly written as
\begin{equation}
  [ \mathbf{M}_{1}, \mathbf{M}_{n} ]+[ \mathbf{M}_{2}, \mathbf{M}_{n-1} ]+\cdots+[ \mathbf{M}_{n-1}, \mathbf{M}_{2} ]+[ \mathbf{M}_{n}, \mathbf{M}_{1} ]=0.
\end{equation}
Furthermore, we can define a formal power series
\begin{equation}
  \mathbf{M}(t) = \sum_{n=1}^{\infty}t^{n-1}\mathbf{M}_{n}
\end{equation}
and the relations are equivalent to
\begin{equation}
  [\mathbf{M}(t), \mathbf{M}(t)] = 0.
\end{equation}

\subsection{Picture assigning procedure}\label{subsec:pcoas}
The method developed in~\cite{Erler2014} is to determine a flow $\mathbf{M}(s, t)$, controlled by an extra variable $s$, on the space of coderivations, such that

1. The $A_{\infty}$ relations are always satisfied,
\begin{equation}\label{eqn:ainf}
[\mathbf{M}(s, t), \mathbf{M}(s, t)] = 0.
\end{equation}

2. One end of the flow $\mathbf{M}(1, 0)$ consists of bosonic string products carrying zero picture number,
\begin{equation}
\mathbf{M}(1, 0) = \sum_{n=1}^{\infty} \mathbf{M}_{n}^{(0)},
\end{equation}
where the superscript denotes the picture number that the product carries,
while on the other end $\mathbf{M}(0, 1)$, every $n$-fold component of the coderivation carries the correct $n-1$ picture number as indicated by the superscripts,
\begin{equation}
\mathbf{M}(0, 1) = \sum_{n=1}^{\infty} \mathbf{M}_{n}^{(n-1)}.
\end{equation}
For future convenience, we also define superscripts in square bracket denoting the picture deficit: for a $n$-product $\mathbf{M}_{n}^{(m)}$ with picture number $m$, the deficit is $n-m-1$ and the product is also written as $\mathbf{M}_{n}^{[n-m-1]}$.

3. All products are in the small Hilbert space.

\noindent The necessity of the last requirement is due to the observation that, to solve for the $A_{\infty}$ relations with picture dressing one needs to introduce the operator $\xi$ that satisfies $\{Q, \xi\} = X$ and carries one picture number.
As mentioned before, $\xi$ can be concretely realized as a line integral of superconformal ghost $\xi(z)$.
However, $\xi$ carries unphysical degrees of freedom.
To eliminate them, one ought to demand that
\begin{equation}
  [\boldsymbol{\eta}, \mathbf{M}] = 0
\end{equation}
as an extra constraint on the flow, here $\boldsymbol{\eta}$ can also be realized as the coderivation from the zero mode of the $\eta(z)$ ghost.
We will see how this requirement, together with the $A_{\infty}$ relation~(\ref{eqn:ainf}), determines the entire flow.
The integrated flow is written as
\begin{equation}
  \begin{split}
    \mathbf{M}(s, t) \equiv\, \sum_{m,n=0}^{\infty}s^{m}t^{n}&\mathbf{M}^{(n)}_{m+n+1} \\
     = Q + s\mathbf{M}_{2}^{(0)} &+ s^{2}\mathbf{M}_{3}^{(0)} + \cdots \\
          + t\mathbf{M}_{2}^{(1)} &+ ts\mathbf{M}_{3}^{(1)} + \cdots \\
                      &+ t^{2}\mathbf{M}_{3}^{(2)} + \cdots,
  \end{split}
\end{equation}
and it satisfies a pair of differential equations
\begin{equation}\label{eqn:m}
  {\frac{\partial} {\partial t} \mathbf{M} ( s, t )}  {{} \,=[ \mathbf{M} ( s, t ), \boldsymbol{\mu} ( s, t ) ],}
\end{equation}
\begin{equation}\label{eqn:mu}
  {\frac{\partial} {\partial s} \mathbf{M} ( s, t )}  {{} \,=[ \boldsymbol{\eta}, \boldsymbol{\mu} ( s, t ) ],}
\end{equation}
where
\begin{equation}
  \boldsymbol{\mu}(s, t) \equiv \sum_{m,n=0} s^{m}t^{n}\boldsymbol{\mu}^{(n+1)}_{m+n+2},
\end{equation}
is an auxiliary coderivation.
It is easy to see that~(\ref{eqn:m}) ensures the $A_{\infty}$ relation on the entire flow by computing
\begin{equation}
\frac{\partial}{\partial t} \left[ \mathbf{M}(s, t), \mathbf{M}(s, t) \right] = \left[ \left[ \mathbf{M}(s, t), \mathbf{M}(s, t) \right], \boldsymbol{\mu}(s, t) \right].
\end{equation}
At $t=0$ the bosonic string products satisfies $\left[ \mathbf{M}(s, 0), \mathbf{M}(s, 0) \right] = 0$ so it is also true for all the other $t$.
On the other hand,~(\ref{eqn:m}) keeps the physical condition, because
\begin{equation}
\frac{\partial}{\partial t} \left[ \boldsymbol{\eta}, \mathbf{M}(s, t) \right] = \left[ \left[\boldsymbol{\eta}, \mathbf{M}(s, t) \right], \boldsymbol{\mu}(s, t) \right] - \frac{1}{2} \left[ \mathbf{M}(s, t), \mathbf{M}(s, t) \right],
\end{equation}
where the second term has been shown to vanish.
We can then conclude that since the bosonic products are in the small Hilbert space $\left[ \boldsymbol{\eta}, \mathbf{M}(s, 0) \right] = 0$, the entire flow is also in the small Hilbert space.

The differential equations~(\ref{eqn:m}) and~(\ref{eqn:mu}) derives a series of recursive relations determining all products.
Recursively we have,
\begin{equation}\label{eqn:mrec}
\mathbf{M}^{(n+1)}_{m+n+2} = \frac{1}{n+1}\sum^{n}_{k=0}\sum^{m}_{l=0}[\mathbf{M}^{(k)}_{k+l+1},\boldsymbol{\mu}^{(n-k+1)}_{m+n-k-l+2}],
\end{equation}
which allows us to compute $\mathbf{M}_{n}^{m+1}$ with $\mathbf{M}_{k}^{l}$ and $\boldsymbol{\mu}_{k}^{l+1}$ for all $0\leq k\leq n,\, 0\leq l \leq m$ at hand.
To solve for~(\ref{eqn:mu}), the auxiliary coderivation $\boldsymbol{\mu}$ is derived from
\begin{equation}\label{eqn:murec}
\mu_{n}^{(m+1)} = (n-1) \xi \circ M_{n}^{(m)},
\end{equation}
which makes sure
\begin{equation}
\left[ \eta, \mu_{n}^{(m+1)} \right] = (n - 1)M_{n}^{(m)}
\end{equation}
and the cyclicity.
Here ``$\xi\circ$'' denotes the operation of taking the average of $\xi$ acting on the output and on each input of a product $b_{n}$, that is
\begin{equation}
\xi\circ b_{n}=\frac{1} {n+1} \left( \xi b_{n}-b_{n} \sum_{k=0}^{n-1} \mathbb{I}^{\otimes k} \otimes\xi \otimes\mathbb{I}^{\otimes n-1-k} \right).
\end{equation}
The average is taken for cyclicity.
Note that acting of $\xi$ introduce an extra picture number to the coderivation and hence the shift of superscript.
Similarly we denote
\begin{equation}
X\circ b_{n}=\frac{1} {n+1} \left( X b_{n}+b_{n} \sum_{k=0}^{n-1} \mathbb{I}^{\otimes k} \otimes X \otimes\mathbb{I}^{\otimes n-1-k} \right)
\end{equation}
for the dressing with line integral of $X(z)$.
To conclude,~(\ref{eqn:mrec}) and~(\ref{eqn:murec}) give a recursive formula for $n$-product $M_{n}^{n-1}$ with the correct $(n-1)$ picture number.

\subsection{\texorpdfstring{$\mathcal{N}=2$}{N = 2} string interaction}

Our desired string products for $ \mathcal{N}=2 $ strings can be built from method in Subsection~\ref{subsec:pcoas} but we have to deal with two picture numbers. 
So we use $(\cdot, \cdot)$ and $[\cdot, \cdot]$ in the superscript to denote the picture number and picture deficit that a product carries.
Unlike the Type II string case where there are worldsheet holomorphic and anti-holomorphic copies $X, \bar{X}$ of the PCOs, $X^{+}$ and $X^{-}$ are not commutative in general.
Therefore, it is not clear how to restore the symmetry between the two picture-changing operators following the construction of~\cite{Erler2014c}, and hence we only consider the asymmetric construction in this paper.
The first step is assigning correct $\pi^{-}$ picture number.
We start with the open bosonic vertices $m_{n}$,
\begin{equation}
  \begin{split}
    m_{1}(\Psi) &= Q\Psi, \\
    m_{2}(\Psi_{1}, \Psi_{2}) &= (-1)^{\mathrm{deg}(\Psi_{1})}\Psi_{1}\ast\Psi_{2}, \\
    m_{n}(\Psi_{1},\dots,\Psi_{n}) &= 0 \quad  \mathrm{for}\,\,\,n\,\geq\,3
  \end{split}
\end{equation}
Consequently the auxiliary products $\mathbf{M}_{n}^{(0, m)}$ and $\boldsymbol\mu_{n}^{(0, m)}$
vanish for $m<n-2$.
\begin{equation}
  \begin{split}
\mathbf{M}^{-}(s^{-}, t^{-}) &\equiv\, \sum_{m=0}^{\infty}\sum_{n=0}^{\infty}(s^{-})^{m}(t^{-})^{n}\mathbf{M}^{(0, n)}_{m+n+1} \\
&=s^{-}\sum_{n=0}^{\infty}(t^{-})^{n}\mathbf{M}^{(0, n)}_{n+2} + \sum_{n=0}^{\infty}t^{n}\mathbf{M}^{(0, n)}_{n+1}.
  \end{split}
\end{equation}
The resulting vertices $M_{n}^{(0, n-1)}$ should be in the small Hilbert space. This condition is written as, in the coalgebra notation,

\begin{equation}\label{eqn:small}
  [\boldsymbol{\eta}^{+}, \mathbf{M}^{-}(s^{-}, t^{-})] = 0,
\end{equation}
where $\mathbf{M}^{-}(t^{-}) = \mathbf{Q} + \sum_{n=1}^{\infty} \mathbf{M}^{(n, 0)}_{n+1}(t^{-})^{n}$ is a formal power series collecting all the vertices with zero $\pi^{+}$ picture number.
Note that the condition $ [\boldsymbol{\eta}^{-}, \mathbf{M}^{-}(s^{-}, t^{-})] = 0 $ automatically holds at this point since we do not introduce any $\xi^{+}$ dressing.
As argued in the last previously, the solution we take satisfies two equations
\begin{equation}
  \frac{\partial}{\partial t^{-}} \mathbf{M}^{-}(s^{-}, t^{-}) = [\mathbf{M}(s^{-}, t^{-}), \boldsymbol{\mu}(s^{-}, t^{-})],
\end{equation}
\begin{equation}
  \frac{\partial} {\partial s^{-}} \mathbf{M}^{-} ( s^{-}, t^{-} ) =[ \boldsymbol{\eta}^{+}, \boldsymbol{\mu}^{-} ( s^{-}, t^{-} ) ]
\end{equation}
Here we introduce auxiliary coderivation flow
\begin{equation}
\boldsymbol{\mu}^{-}(s^{-}, t^{-}) = \sum_{n=0}^{\infty} \boldsymbol{\mu}^{(0,n+1)}_{n+2}(t^{-})^{n}.
\end{equation}
to facilitate the small Hilbert space condition~(\ref{eqn:small}).
Note that equations are simplified since $\mathbf{M}^{-}(s^{-}, t^{-})$ contains at most linear terms in $s^{-}$.

The next step is to assgin the $\pi^{+}$ picutre number, now we have to start with a set of ``bosonic'' $\mathbf{M}^{-}(0, 1)$ vertices with non-vanishing higher-point parts.
We again want to determine a flow
\begin{equation}
  \mathbf{M}(s^{+}, t^{+}) = \sum_{m=0}^{\infty}\sum_{n=0}^{\infty}\mathbf{M}^{(n,m+n)}_{m+n+1}(s^{+})^{m}(t^{+})^{n}
\end{equation}
and set $\mathbf{M}(1, 0) = \mathbf{M}^{-}(0, 1)$.
We have the differential equations
\begin{equation}
  \frac{\partial}{\partial t^{+}} \mathbf{M}(s^{+}, t^{+}) = [\mathbf{M}(s^{+}, t^{+}), \boldsymbol{\mu}(s^{+}, t^{+})],
\end{equation}
\begin{equation}
  \frac{\partial} {\partial s^{+}} \mathbf{M} ( s^{+}, t^{+} ) =[ \boldsymbol{\eta}^{-}, \boldsymbol{\mu} ( s^{+}, t^{+} ) ]
\end{equation}
where
\begin{equation}
  \boldsymbol{\mu}(s^{+}, t^{+}) = \sum_{m=0}^{\infty}\sum_{n=0}^{\infty}\boldsymbol{\mu}^{(m,n+1)}_{n+2}(s^{+})^{m}(t^{+})^{n}.
\end{equation}

The four differential equations derives a series of recursive relations determining all products.
From the bosonic string product
\begin{equation}
\mathbf{M}^{-}(1, 0) = \mathbf{Q} + \mathbf{m}_{2}
\end{equation}
we have, recursively
\begin{equation}
\mathbf{M}^{(0, n+1)}_{n+2} = \frac{1}{n+1}\sum^{n}_{m=0}[\mathbf{M}^{(0, m)}_{m+1},\boldsymbol{\mu}^{(0, n-m+1)}_{n-m+2}],
\end{equation}
\begin{equation}
\mathbf{M}^{(0, n)}_{n+2} = \frac{1}{n}\sum^{n-1}_{m=0}[\mathbf{M}^{(0, m)}_{m+2},\boldsymbol{\mu}^{(0, n-m)}_{n-m+1}],
\end{equation}
and
\begin{equation}
\mu^{(0,n+1)}_{n+2} = \xi^{+} \circ M^{(0,n)}_{n+2}.
\end{equation}
All the auxiliary products $\mathbf{M}_{n}^{(0,m)}$ and $\boldsymbol{\mu}_{n}^{(0,m)}$ vanish for $m < n-2$.

For the $ \pi^{+} $ picture number iteration, we have
\begin{equation}
\mathbf{M}^{(n+1,m+n+1)}_{m+n+2} = \frac{1}{n+1} \sum^{n}_{k=0}\sum^{m}_{l=0} \left[ \mathbf{M}^{(k,k+l)}_{k+l+1}, \boldsymbol\mu^{(n-k+1,m+n-k-l+1)}_{m+n-k-l+2} \right]
\end{equation}
and
\begin{equation}
\mu^{(m+1,m+n+1)}_{m+n+2} = (n+1)\xi^{-} \circ M^{(m,m+n+1)}_{m+n+2}.
\end{equation}

\section{Scattering amplitudes}
\label{sec:amp}

In this section we show that the action we just constructed algebraicly from gauge invariance gives the correct string amplitudes.
The strategy is similar to \cite{Konopka2015}, where the flow on the space of coderivations allows us to move all the PCOs to the external legs with on-shell vanishing corrections.
We let $\mathcal{S}(\mathbf{M})$ denote the minimal model of coderivation $\mathbf{M}$ (possibly a flow with formal variables), which is related to the scattering matrix $S(\mathbf{M})$ of the physical theory as
\begin{equation}\label{eqn:smatrix}
  S(\mathbf{M}) = \omega_{L}(\mathbb{I}\otimes\xi^{+}\xi^{-}P_{1} \mathcal{S}(\mathbf{M})).
\end{equation}
The scattering matrix is defined on the on-shell states in the cohomology of $Q$.
If we consider the projection $P$ to $\mathrm{ker}(Q)$, we can construct a contracting homotopy $H$, such that
\begin{equation}
  \begin{split}
    Q H + H Q &= \mathbb{I} - P,\\
    H P &= 0,\\
    P H &= 0,\\
    H^{2} &= 0.
  \end{split}
\end{equation}
We promote $P, H$ to the level of derivation $\mathbf{P}, \mathbf{H}$.
$\mathcal{S}(\mathbf{M})$ can be computed using homological perturbation theory~\cite{Konopka2015},
\begin{equation}
  \mathbf{P}\Big(\mathbf{Q}+\mathbf{M}_{\mathrm{int}}(1-\mathbf{H}\mathbf{M}_{\mathrm{int}})^{-1}\Big)\mathbf{P},
\end{equation}
where $\mathbf{M}_{\mathrm{int}} = \mathbf{M} - \mathbf{Q}$ is the interacting part.
To recursively check the scattering amplitude relation, we can compute that partial derivatives
\begin{equation}
    \frac{\partial}{\partial s^{+}} \mathcal{S}(\mathbf{M}) = [\boldsymbol{\eta}^{-}, \hat{P}(1-\mathbf{M}_{\mathrm{int}}H)^{-1}\boldsymbol{\mu}(1 - H\mathbf{M}_{\mathrm{int}})^{-1}\hat{P}] \equiv [\boldsymbol{\eta}^{-}, \boldsymbol{\rho}]
\end{equation}

The right hand side is $\eta^{-}$-exact.
Since $\xi^{+}\circ$ is a contracting homotopy for $\eta^{-}$, we can decompose $\rho$ into $\xi^{+} \circ$ part and $\eta^{-}$-exact part,
\begin{equation}\label{eqn:etacoh}
\xi^{+} \circ \frac{\partial}{\partial s^{+}} \mathcal{S}(\mathbf{M}) = \boldsymbol{\rho} - [\boldsymbol{\eta}^{-}, \xi^{+} \circ \boldsymbol{\rho}]
\end{equation}
To get the desired PCO insertion, we take the BRST derivation of~(\ref{eqn:etacoh})
\begin{equation}
\begin{split}
X^{+} \circ \frac{\partial}{\partial s^{+}} \mathcal{S}(\mathbf{M}) = \left[ \mathbf{Q}, \boldsymbol{\rho} \right] - \Big[ \mathbf{Q}, [\boldsymbol{\eta}^{-}, \xi^{+} \circ \boldsymbol{\rho}] \Big].
\end{split}
\end{equation}
Note that
\begin{equation}
[\mathbf{Q}, \boldsymbol{\rho}] =
 \hat{P}(1-\mathbf{M}_{\mathrm{int}}H)^{-1}[\mathbf{M}, \boldsymbol{\mu}](1 - H\mathbf{M}_{\mathrm{int}})^{-1}\hat{P} = \frac{\partial}{\partial t^{+}} \mathcal{S}(\mathbf{M}),
\end{equation}
which is just
\begin{equation}
   \begin{split}
     &\frac{\partial}{\partial t^{+}} \mathcal{S}(\mathbf{M}) \\
     &= \hat{P}(1-\mathbf{M}_{\mathrm{int}}\mathbf{H})^{-1}\frac{\partial}{\partial t^{+}}\mathbf{M}\mathbf{H}(1 - \mathbf{H}\mathbf{M}_{\mathrm{int}})^{-1}\mathbf{M}_{int}\hat{P} + \hat{P}(1-\mathbf{M}_{\mathrm{int}}\mathbf{H})^{-1}\frac{\partial}{\partial t^{+}}\mathbf{M}\hat{P} \\
     &= \hat{P}(1-\mathbf{M}_{\mathrm{int}}H)^{-1}[\mathbf{M}, \boldsymbol{\mu}](1 - H\mathbf{M}_{\mathrm{int}})^{-1}\hat{P}.
   \end{split}
 \end{equation}
So we have
\begin{equation}
  X^{+}\circ \frac{\partial}{\partial s^{+}} \mathcal{S}(\mathbf{M}) = \frac{\partial}{\partial t^{+}} \mathcal{S}(\mathbf{M}) - [\mathbf{Q}, [\boldsymbol{\eta}^{-}, \xi^{+} \circ \boldsymbol{\rho}]] = \frac{\partial}{\partial t^{+}} \mathcal{S}(\mathbf{M}) - [\boldsymbol{\eta^{-}}, [\mathbf{Q}, \mathbf{T}]],
\end{equation}
where $\mathbf{T}$ is some irrelavent part, since the $\xi^{+}$ in~(\ref{eqn:smatrix}) strips off $\mathbf{\eta}^{-}$, leaving only $Q$-exact term.
This means that up to irrelavent part we have
\begin{equation}
  \mathcal{S}(\mathbf{M})^{[m,0]}_{n} = \frac{m+1}{n-m-1}X^{+}\circ\mathcal{S}(\mathbf{M})_{n}^{[m+1,0]},
\end{equation}
where the square bracket notation denotes the picture deficits.
Specifically here it means that $\mathcal{S}(\mathbf{M})^{[m,0]}_{n}$ is the coefficient of $(s^{+})^{m}(t^{+})^{n}$.
After reduction we have
\begin{equation}
\mathcal{S}(\mathbf{M})^{[0,0]}_{n} = (X^{+})^{n-1}\circ\mathcal{S}(\mathbf{M})^{[n-1,0]}_{n} = (X^{+})^{n-1}\circ\mathcal{S}(\mathbf{M}^{-}(0, 1)),
\end{equation}
that is to say
\begin{equation}
\mathcal{S}(\mathbf{M}(1,0)) = (X^{+})^{n-1}\circ\mathcal{S}(\mathbf{M}^{-}(0, 1)).
\end{equation}
Similarly for the flow $\mathbf{M}^{-}$ we have
\begin{equation}
  X^{-}\circ \frac{\partial}{\partial s^{-}} \mathcal{S}(\mathbf{M}^{-}) = \frac{\partial}{\partial t^{-}} \mathcal{S}(\mathbf{M}^{-}) - [\boldsymbol{\eta^{+}}, [\mathbf{Q}, \mathbf{T}^{-}]],
\end{equation}
yielding
\begin{equation}
\mathcal{S}(\mathbf{M}^{-})^{[0,0]}_{n} = (X^{-})^{n-1}\circ\mathcal{S}(\mathbf{M}^{-})^{[n-1,n-1]}_{n},
\end{equation}
namely
\begin{equation}
\mathcal{S}(\mathbf{M}^{-}(1, 0)) = (X^{-})^{n-1}\circ\mathcal{S}(\mathbf{M}^{-}(0, 1)) = (X^{-})^{n-1}\circ\mathcal{S}(\mathbf{Q} + \mathbf{m}_{2})
\end{equation}

In a word, the action we constructed has a S-matrix of
\begin{equation}
  \begin{split}
    S_{n+1}(\mathbf{M}(1, 0)) &= \omega_{L}(\mathbb{I} \otimes \xi^{+}\xi^{-}P_{1}\mathcal{S}(\mathbf{M}(1, 0))) \\
    &= \omega_{L}\Big(\mathbb{I} \otimes \xi^{+}\xi^{-}P_{1}(X^{+})^{n-1}\circ(X^{-})^{n-1}\circ\mathcal{S}(\mathbf{Q}+\mathbf{m}_{2})\Big) \\
  \end{split}
\end{equation}
Since moving the PCOs around only results in BRST-exact term, we conclude that
\begin{equation}
S_{n+1} = S^{\mathrm{bosonic}}_{n+1}(X^{+})^{n-2}(X^{-})^{n-2},
\end{equation}
where the $X^{\pm}$ insertions are arbitrary on the worldsheet, so we reproduce the worldsheet string amplitudes in~(\ref{eqn:amp}) at $n_{I}=0$.

\section{Summary and discussion}%
\label{sec:dis}
We have constructed a string field theory action with cyclic $A_{\infty}$ structure for the open $\mathcal{N} = 2$ string.
We argued that, by restrcting the string field to the kernel of $\tilde{b}_{0} = 0$, we can write the free theory in the relative BRST cohomology sense.
Furthermore, the spectral flow induced isomorphism allows us to consider the Neveu-Schwarz sector only, thereby simplifying the theory.
The construction of interacting vertices proceeds via a two-step procedure that successively assigns both the correct $\pi^{-}$ and $\pi^{+}$ picture number in turn.
Moreover, the S-matrix of this theory has been show to reproduce those obtained from the worldsheet approach.

It would be a straightforward generalization to include the worldsheet antiholomorphic sector of closed string.
This construction is purely algebraic and it is interesting to explore its relation to the $\mathcal{N} = 2$ supermoduli space.
More curious is the fact that the $n$-point worldsheet amplitudes of genus $g$ of $\mathcal{N} = 2$ string have been shown to vanish for $n>3$ and $g>0$~\cite{BerkovitsTop1995}.
This suggests that in principle one can find a quasi-isomorphic loop homotopy algebra truncated at cubic level, and the cubic vertex alone suffices to cover the entire $\mathcal{N} = 2$ supermoduli space, similar to Witten's bosonic theory.
We hope the method developed in~\cite{Ohmori2018} can provide further insights into this problem.
Moreover, a cubic action may initialize a generalization of recently realized homotopy double copy~\cite{Borsten2021} at the string level.

Our construction focuses on the zero-instanton sector just like Berkovits formulation.
It has been shown that, rather surprisingly, including all instanton sectors does not change the dynamics but only affects gauge~\cite{Lechtenfeld1997}, and the field theory amplitudes can be derived from a cubic self-dual Yang-Mills action.

Finally, the connection between the homotopy-algebraic action and Berkovits' WZW-like action is well-established in the $\mathcal{N} = 1$ case~\cite{Erler2015}, but remains largely unexplored for $\mathcal{N} = 2$ string theory.
We hope to come to these issues in future work.

\acknowledgments%
The author thanks Yuji Okawa for helpful discussions and Masashi Hamanaka for supervision during his PhD program and for introducing him the topic of $\mathcal{N} = 2$ string theory.
This work was financially supported by JST SPRING, Grant Number JPMJSP2125.
The author would like to thank the “THERS Make New Standards Program for the Next Generation Researchers”.

\bibliography{main}

\end{document}